\documentclass{article}
\usepackage{spconf,graphicx,subfig}
\usepackage[cmex10]{amsmath}
\usepackage{amssymb}
\usepackage{soul}
\usepackage{xcolor}
\usepackage{textcomp}
\def\L{{\cal L}}

\def\z{\boldsymbol{z}}

\def\A{\mathcal{A}}

\def\bAhat{\mathbf{\hat{A}}}

\def\etal{et al.\ }
\def\th{\mathrm{th}}

\def \obj {\boldsymbol{f}}

\def \L {\mathcal{L}}

\def \P {\mathbb{P}}
\def \path{\mathbb{P}}

\def \f{\boldsymbol{f}}

\def \pr{\mathrm{pr}}
\def \Pr{\mathrm{Pr}}

\title{A list-mode OSEM-based attenuation and scatter compensation method for SPECT}

\name{Md Ashequr Rahman$^{1,2}$ \quad Richard Laforest${^2}$ \quad Abhinav~K.~Jha$^{1,2}$\thanks{This work was financially supported by NIH R21 EB024647 (Trailblazer award) and an NVIDIA GPU grant. }}

\address{$^1$Department of Biomedical Engineering, Washington University in St. Louis, St. Louis, MO, USA\\
$^2$Mallinckrodt Institute of Radiology, Washington University in St. Louis, St. Louis, MO, USA}
\begin{document}
%\ninept
%

\twocolumn[
  \begin{@twocolumnfalse}
  \section*{IEEE Copyright Notice:}
   \textcopyright 2020 IEEE.  Personal use of this material is permitted.  Permission from IEEE must be obtained for all other uses, in any current or future media, including reprinting/republishing this material for advertising or promotional purposes, creating new collective works, for resale or redistribution to servers or lists, or reuse of any copyrighted component of this work in other works.
   
   \vspace{100pt}
   Accepted to be published in: 2020 IEEE International Symposium on Biomedical Imaging (IEEE ISBI 2020), April 3-7, 2020
  \end{@twocolumnfalse}
  ]

\clearpage
\maketitle
\begin{abstract}
Reliable attenuation and scatter compensation (ASC) is a pre-requisite for quantification and beneficial for visual interpretation tasks in SPECT. In this paper, we develop a reconstruction method that uses the entire SPECT emission data, i.e. data in both the photopeak and scatter windows, acquired in list-mode format and including the energy attribute of the detected photon, to perform ASC. We implemented a GPU-based version of this method using an ordered subsets expectation maximization (OSEM) algorithm. The method was objectively evaluated using realistic simulation studies on the task of estimating uptake in the striatal regions of the brain in a 2-D dopamine transporter (DaT)-scan SPECT study. 
We observed that inclusion of data from the scatter window and using list-mode data yielded improved quantification compared to using data only from the photopeak window or using binned data. These results motivate further development of list-mode-based ASC methods that include scatter-window data for SPECT.
%\hlc[cyan]{The method yielded improved and reliable quantitative performance for list-mode data encompassing the full energy spectrum compared to list-mode data within partial energy spectrum or data processed in binned mode.}
\end{abstract}
\begin{keywords}
SPECT, Reconstruction, Attenuation and scatter compensation, List-mode data, Brain, Parkinson's disease.
\end{keywords}
\section{Introduction}
\label{sec:intro}
Single-photon emission computed tomography (SPECT) has an important role in the diagnosis and therapy of several diseases such as Parkinson's disease, coronary artery disease, and many cancers. A major image-degrading process in SPECT is the scatter and resultant attenuation of photons as they traverse through the tissue before they reach the detector. Reliable attenuation and scatter compensation (ASC) is a pre-requisite for quantification tasks, such as quantifying biomarkers from SPECT images \cite{bailey2013evidence} or performing SPECT-based dosimetry \cite{song2011development,jha2016no}. Also, ASC has been observed to be beneficial for visual interpretation tasks \cite{garcia2007spect}. Thus, there is an important need for reliable ASC methods.

In this paper, we focus on developing ASC methods when an attenuation map is available from a transmission scan, typically a CT scan.
Several such methods have been developed \cite{hutton2011review,ljungberg2018absolute,king1995attenuation, hashimoto1997scatter}. 
However, typically existing methods do not use the precise value of the energy attribute of the detected photon, as is available when the data are stored in list-mode (LM) format.
%In spite of these tremendous advances, there remains a need for improved methods for ASC. 
Using this precise value provides an avenue to improve the ASC. 
This was a major motivation for ASC approaches based on extensive spectral analysis and modeling \cite{mas1990scatter,devito1989energy}. 
Prior work in PET imaging has shown that using LM data and incorporating energy information led to improved ASC \cite{guerin2010novel,popescu2006pet}. More recently, we observed that SPECT LM emission data containing the energy attribute contains information to jointly estimate the activity and attenuation distributions \cite{rahman2019fisher}.
Also, several studies have shown that LM data can yield improved reconstruction and quantification compared to binned data in SPECT imaging \cite{Bouwens:2001, caucci2019towards,jha2015estimating, jha2015singular,Jha:13:Fully3d, Henscheid:2017}.
These investigations motivate the use of SPECT LM data containing the energy attribute for ASC.

Existing SPECT reconstruction methods also typically use only the photo-peak (PP) window data for estimating the activity. In this context, recent studies have shown that addition of data from scatter window can provide 
more information to estimate the activity uptake compared to using data from only the PP window \cite{rahman2019fisher,kadrmas1997analysis}. 
Using data from PP and scatter windows for activity estimation also increases the effective sensitivity of SPECT systems, that otherwise, is typically very low ($<100$ counts/million emitted counts). Based on these scientific premise, we hypothesize that processing the entire SPECT data from PP and scatter windows in LM format containing the energy attribute can provide improved quantification compared to using binned data or using only PP window data.

To investigate this hypothesis, we developed a method to perform ASC using the SPECT emission data acquired in LM format and containing the energy attribute. The method is inspired by the LM reconstruction approach for PET imaging proposed by Parra \etal \cite{parra1998list} and extends upon theory originally briefly proposed by Jha \etal for jointly estimating the activity and attenuation distribution from SPECT LM data \cite{jha2013joint}. We developed this theory specifically for estimating the activity distribution with known attenuation and derived a maximum-likelihood expectation-maximization (MLEM) algorithm for this task.  
An ordered-subsets version of this algorithm was then developed, and implemented on GPU-based hardware for computational efficiency.
The method was objectively evaluated using simulation studies in the context of a quantitative 2D dopamine transporter (DaT)-scan SPECT study.
\begin{figure}
\centering
\includegraphics[height = 2 in]{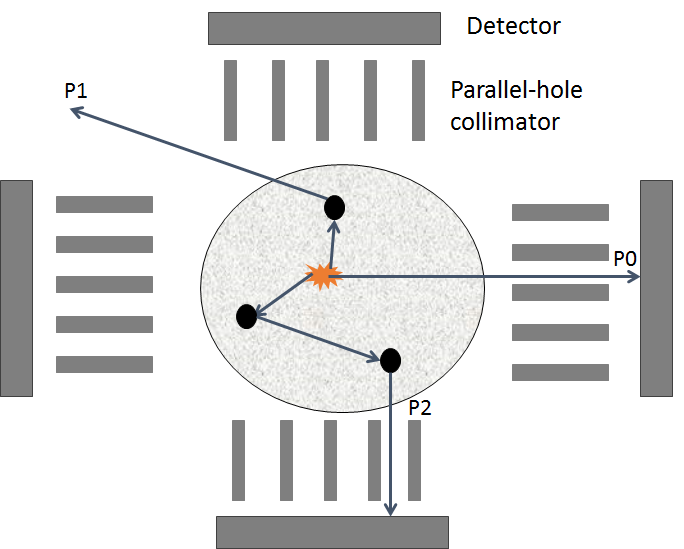}
\caption{A schematic of SPECT system demosnstrating the definition of a path.}
\label{fig:spect_system}
\end{figure}
\section{Methods}\label{theory}
\subsection{THEORY}
Consider a preset-time scintillation-detector-based SPECT  system imaging an activity distribution denoted by the vector $\f$. The system acquires and stores data in LM format over  a fixed acquisition time, $T$. Let $J$ denote the number of detected events. Note that the proposed technique is also applicable to a preset-count system. Denote attributes collected for the $j^{\th}$ LM event by the attribute vector $\bAhat_j$. This vector contains attributes such as the position of interaction with the scintillator, energy deposited in the scintillator, time of interaction, and the angular orientation of the detector that interacted with the photon. Denote the full LM dataset as a set of attribute vector $\hat{\A}=\{\bAhat_j,j=1,2,...J\}$. Since the detected LM events are independent, the LM data likelihood is given by
\begin{align}
\pr(\hat{\A},J|\obj) = \Pr(J|\obj)\prod_{j=1}^{J}\pr(\bAhat_j|\obj),
\label{eq:lm_ll_gen}
\end{align} 
\noindent Our approach to develop the reconstruction technique is to estimate $\obj$ that maximizes the probability of $\mathcal{\hat{A}}$. While obtaining the expression for $\Pr(J | \f)$ is easy since $J$ is Poisson distributed, deriving an expression for $\pr(\bAhat_j| \f)$ is complicated. To address this issue, we use the fact that each detected photon traverses through a specific discrete sub-unit of space after being emitted. We refer to this sub-unit as a path \cite{rahman2019fisher,jha2013joint}. For example, in Fig.\ref{fig:spect_system}, P1, P2, and P3 denote 3 such paths.  Note that we do not know in advance of the path that a photon takes. To address this issue while deriving our reconstruction method, we define a latent variable $z_{j,\P}$ as follows:
\[
 z_{j,\P} = 
  \begin{cases} 
   1 & \text{if event $j$ took the path $\P$.} \\
   0       & \text{otherwise}.
  \end{cases}
\]

\noindent Defining this latent variable enables developing an expectation-maximization (EM) technique to perform the reconstruction. Further, it enables deriving the expression for $\pr(\bAhat_j| \path)$. More specifically, we can expand the term in (\ref{eq:lm_ll_gen}) in terms of a mixture model as
\begin{equation}
\pr(\bAhat_j|\obj)=\sum_{\P}\pr(\bAhat_j|\P)\Pr(\P|\obj).
\label{eq:Aj_expansion}
\end{equation} 
The number of detected events $J$ is Poisson distributed with mean $\beta T$ where $\beta$ is the mean rate of detected photons. Using this fact and starting from Eq.~\ref{eq:lm_ll_gen}, we can write the log-likelihood of the acquired LM data, denoted by $\L(\obj|\hat{\A},J)$ as 
\begin{align}
\L(\obj|\hat{\A},J)=\sum_{j=1}^{J}\log\sum_{\P}\pr(\bAhat_j|\P)\Pr(\P|\obj) +\nonumber\\
J\log(\beta T)-\beta T - J!.\label{eq:lll_med}
\end{align}
The term $\Pr(\P | \f)$ represents the radiation that is transmitted through the path $\P$. The expression for this term is given by \cite{rahman2019fisher}
\begin{align}
\pr(\P|\obj)=\frac{\obj({\P})s_{eff}(\P)}{\sum_{\P'}\obj({\P'})s_{eff}(\P')},
\label{eq:pr_path_1}
\end{align}   
where $\obj(\P)$ denotes the activity of the starting voxel of the path, $\P$. The term $s_{eff}(\P)$ is independent of object activity, and models the sensitivity of the path to the detector surface. This term models the attenuation and scatter of photons in the tissue as well as the transmission of photons through the collimator. Using these expressions, we can derive the log-likelihood of the observed LM data and the latent variables, denoted by $\mathcal{L}_C$, as 
\begin{align}
&\L_C(\obj|\{\bAhat_j,\z_j\}_{j=1,2...J},J)\nonumber\\
&=\sum_{j=1}^{J}\left[\sum_{\P}z_{j,\P}\left\{\log \pr(\bAhat_j|\P)+\log \obj({\P}) + \log s_{eff}(\P)\right\}\right] \nonumber\\
& \quad\quad -T\sum_{\P}\obj({\P})s_{eff}(\P).
\label{eq:cd_lll}
\end{align}
Having defined the log-likelihood, we put forth the LM MLEM (LM-MLEM) technique. In the expectation (E) step of iteration $(t+1)$, we take the expectation of~(\ref{eq:cd_lll}) conditioned on observed data using the previous estimate of object activity distribution, $\f^{(t)}$. The result is equivalent to replacing $z_{j,\P}$ in (\ref{eq:cd_lll}) with its expected value conditioned on the observed data. 
In the maximization (M) step, we maximize this conditional expectation of the log likelihood. This yields the following iterative update equation:
\begin{align}
f_q^{(t+1)}=\dfrac{\sum_{j=1}^{J}\sum_{\P_q}\bar{z}_{j,\P_q}^{(t+1)}}{T\sum_{\P_q}s_{eff}(\P_q)},
\label{eq:lm_mlem}
\end{align}
where $f_q$ and $\P_q$ denote the activity in the $q^{th}$ voxel and the paths that originate from the $q^{th}$ voxel, respectively. 
\subsection{THE ORDERED-SUBSET LM-MLEM (OS-LM-MLEM)}\label{sec:os-lm-mlem}
The proposed method is computationally expensive. To reduce the compute time, we developed an ordered-subsets (OS) version of the technique, and implemented this version on parallelized computing hardware. We first describe the developed OS version.

The technique of OS is widely used to achieve faster convergence of EM-based reconstruction methods \cite{hudson1994accelerated}.
To develop the OS version of the developed LM-MLEM technique, similar to conventional OSEM-based algorithms, we divided the LM data into subsets based on the detector angle of each LM event. In each sub-iteration, an estimate of the activity uptake is reconstructed using all the events in a subset. Denote the subset by the index $i$. Also, let $S_i^{\P}$ and $S_i^{j}$ denote the set of paths that reach the detector and the set of events that are detected at angles that are elements of the subset $i$, respectively. Then, starting from~\eqref{eq:lm_mlem}, the iterative update corresponding to the 
$i$'th subset and $(t+1)$'th iteration is derived to be 
\begin{align}
f_q^{(t+1,i)}=\dfrac{\sum\limits_{\substack{j=1\\j\epsilon S^j_i}}^{J}\sum\limits_{\P_q\epsilon S^{\P}_i}\bar{z}_{j,\P_q}^{(t+1,i)}}{T\sum\limits_{\P_q\epsilon S_i^{\P}}s_{eff}(\P_q)}&;i=1,2..,N_s,\nonumber\\
&,t=0,1,...,N_g-1,
\label{eq:os-lm-mlem}
\end{align}
where $N_g$ denote the number of global iterations and $N_s$ is the number of sub-iterations or equivalently the number of subsets. $\bar{z}_{j,\P_q}^{(t+1,i)}$ in (\ref{eq:os-lm-mlem}) can be computed as follows:
\begin{align}
\bar{z}_{j,\P_q}^{(t+1,i)}=\dfrac{\pr(\bAhat_j|\P_q)Pr(\P_q|\f^{(\boldsymbol{\eta})})}{\sum_{\P'}Pr(\bAhat_j|\P')\pr(\P'|\f^{(\boldsymbol{\eta})})},\label{eq:z_bar_update_os}
\end{align} 
where $\boldsymbol{\eta}$ is a $2$-D vector denoting the iteration state given by 
\begin{align}
\boldsymbol{\eta} = \begin{cases}
(t,N_s) &,\text{if } i=0.\\
(t+1,i-1) &,\text{otherwise.}
\end{cases}
\end{align}

At any iteration, the number of calculation scales linearly as number of non-zero voxels, $N_{nz}$. 
The reconstruction time was reduced by mimimizing $N_{nz}$ in the initial estimate of the activity map.
This was done by generating an initial crude estimate of phantom boundary using OSEM reconstruction from binned sinogram. 
Further, the computational complexity increases exponentially as the order of scatter. To reduce the number of paths, we considered only up to first-order scatter events and all the scatter was assumed to occur in plane. 

To further increase the computational speed up, the method was parallelized and implemented on NVIDIA GPU hardware. In each iteration of the OSEM, the evaluation of the denominator of \eqref{eq:z_bar_update_os} and the numerator of \eqref{eq:os-lm-mlem} for a fixed voxel $q$  were performed in parallel using reduction algorithm. 

\section{Objective evaluation of proposed technique}
The proposed method was evaluated in the context of a  quantitative 2-D DaT-scan SPECT study, where the task was to estimate the mean activity uptake in the caudate and putamen regions of the reconstructed SPECT image. There is much interest in exploring whether uptakes in these regions can help with improved diagnosis of Parkinson's disease. 
%To evaluate the performance of the proposed method, we conducted a realistic SPECT simulation study. 
A 2D clinical SPECT scanner with a geometry similar to the Optima 640 parallel-hole collimator and imaging uptake of Ioflupane (I-131) tracer within the brain was simulated \cite{rahman2019fisher}. Patient anatomy and physiology were modeled using the Zubal digital brain phantom \cite{zubal1994computerized}. The LM data acquisition was modeled using a Monte Carlo-based code, where for each photon, we collected the position of interaction, energy of the photon, and the angular orientation of the detector. To simulate a low-dose setting, we collected around one-third of the number of photons typically acquired clinically. 

The proposed OS-LM-MLEM reconstruction method was used to estimate the activity map. 
%All reconstructions were done on a system consisting of four NVIDIA V100 GPUs and one Intel Xeon CPU. From the reconstructed images, we estimated the total activity in the caudate and putamen regions. 
The experiments were repeated for multiple noise realizations. The normalized root-mean-square error (RMSE) of the estimated activity uptakes were computed.  

We evaluated the effect of including data from the scatter window on estimating the activity map . For this purpose, we considered three configurations, namely, using data only from PP window, using data only from photons with energy higher than 120 keV, and using data from entire energy spectrum.

We also evaluated the effect of binning the energy and position attributes of the LM data on quantification performance. 
The position attribute was binned into $64$ bins and the energy attribute into 2 and 3 bins in different experiments.
%We also reconstructed the activity map using the more conventional triple-energy-window (TEW)-based \cite{ogawa1991practical} ASC method for comparison. For TEW, we used a $20\%$ energy window, centered at the photo-peak energy, for the main window and two overlapping $7$ KeV scatter window on each side of the main window. 
\begin{figure}[t!]
\centering
\subfloat[]{
\includegraphics[height = 1.1in]{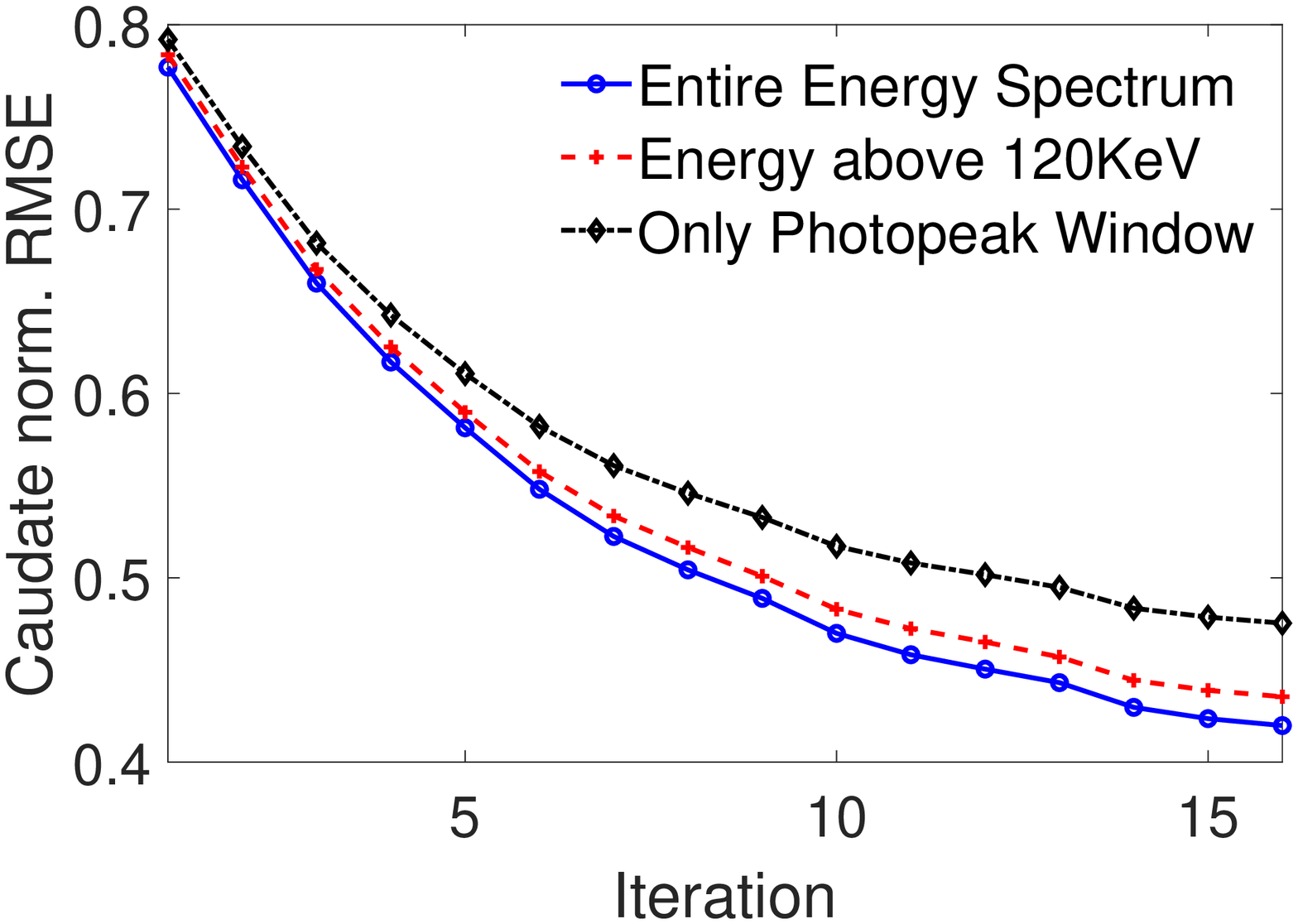}
\label{fig:cd_bias}
}
\subfloat[]{
\includegraphics[height = 1.1in]{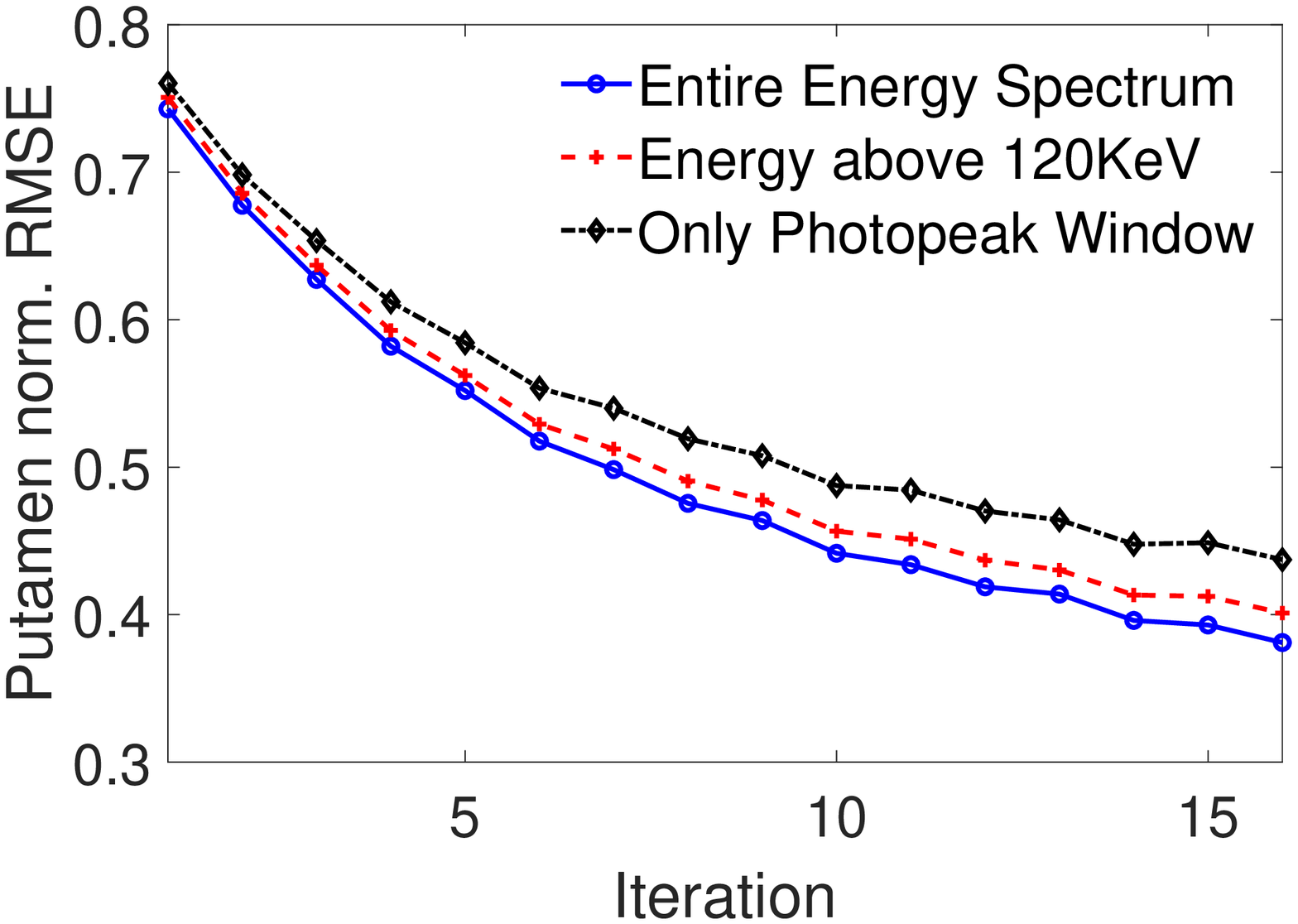}
\label{fig:cd_mse}
}
\hfill
\subfloat[]{
\includegraphics[height = 1.1in]{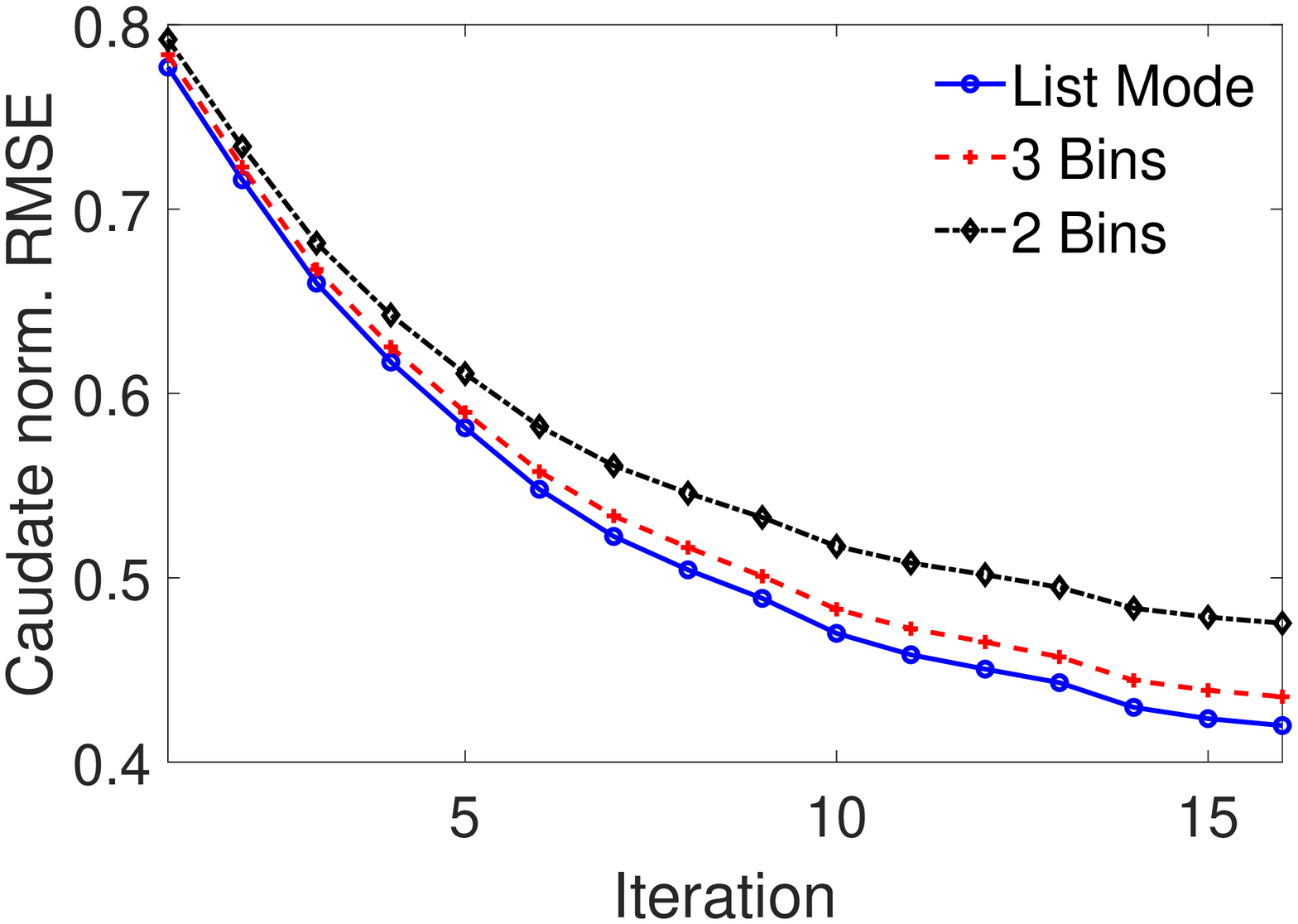}
\label{fig:pt_bias}
}
\subfloat[]{
\includegraphics[height = 1.1in]{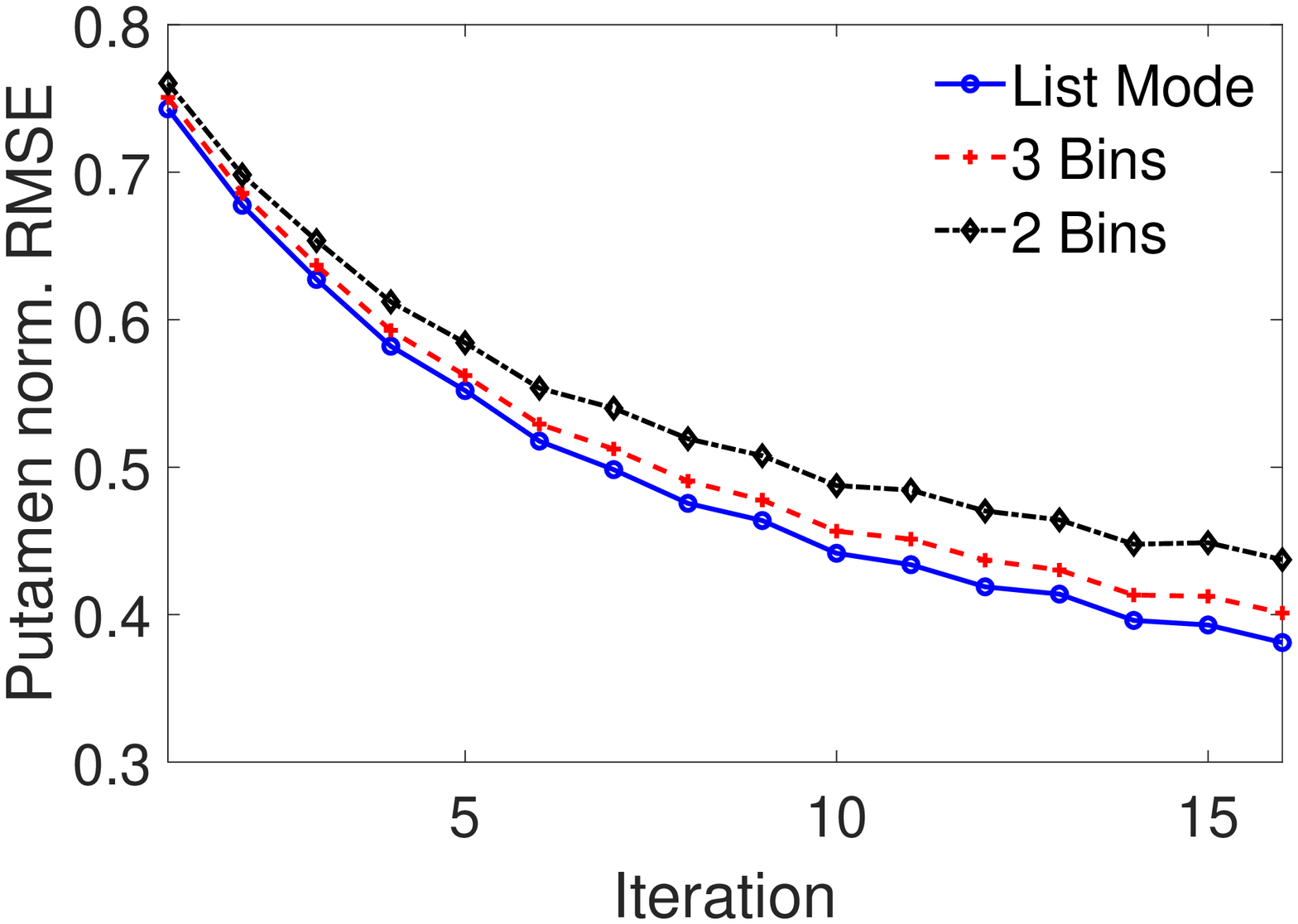}
\label{fig:pt_mse}
}

\caption{Normalized RMSE plot as a function of iteration number for (a,b) different configurations of energy window and (c,d) different number of enrgy bins. The left and right side plots are for caudate and putamen, respectively. }
\label{fig:bvm}
\end{figure}

\begin{figure}[t!]
\centering

\includegraphics[height = 2.4in]{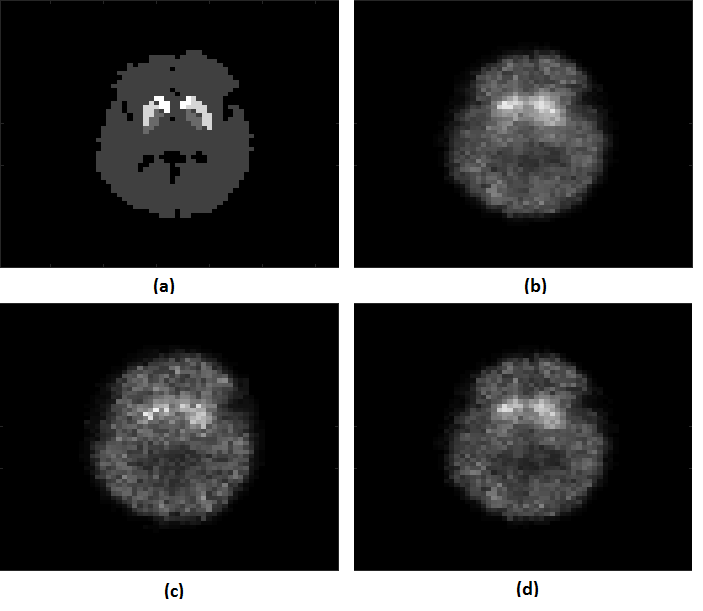}
\label{fig:br_act_true}

\caption{(a) True activity and reconstructed image using OS-LM-MLEM (b) with LM data and the entire energy spectrum; (c) with the entire energy spectrum but binned data with 2 energy bins; (d) with LM data only from PP window. The reconstructed images (b,c,d) are represented on the same scale. }
\label{fig:recon}
\end{figure}

\section{Results}\label{sec:result}
In Fig.~2, the normalized RMSE plots for the activity uptake in the caudate and putamen regions are shown as a function of iteration for the different configurations. Figs.~2a and 2b show that as we increased the range of energies considered, the RMSE reduced.
For example, including data from the entire energy spectrum resulted in approximately $5\%$ decrease in the RMSE for activity uptake in both caudate and putamen compared to using only data from PP window.
Figs.~2c and 2d show that using LM data yielded lower RMSE for both caudate and putamen regions. 
For example, when using LM data, the RMSE of activity uptake decreased by $6\%$ in caudate and $5\%$ in putamen compared to using two energy bins.
Representative reconstructed images with different configurations are shown in Fig.~3. Visually also, the results with LM data and using the entire energy spectrum appear to have improved quality.
Overall, these results demonstrate that data processed in LM format and encompassing all the emission and scattered photons yielded superior performance.

A pure LM-MLEM based technique was also developed and compared to the OS-LM-MLEM technique. We found that the reconstruction results were similar and as expected, the OSEM technique with four subsets yielded close to four-times computational speedup. 
%p_e2_43.36
%p_e3_41.23
%p_e0_38.11
%c_e2_48.29
%c_e3_45.2
%c_e0_41.98
%p_t143_43.73
%p_t120_40.11
%p_tn_38.11
%c_t143_47.54
%c_t120_43.55
%c_tn_41.98
\section{Conclusions and Future Work}\label{sec:disc}
In this manuscript, we proposed an OS-LM-MLEM-based reconstruction method that uses both the scattered and photopeak data acquired in LM format to perform ASC in SPECT. Results from realistic simulation studies conducted in the context of measuring regional activity uptakes in a 2-D DaT-scan SPECT study demonstrated that inclusion of data in the scatter window yielded improved quantification compared to using only PP data. Further, processing data in LM format yielded improved quantification compared to binning the energy and position attributes. 
A challenge with the proposed method is the large computational time. To address this, improved optimization strategies will need to be developed. 
Additionally, further development of the method for 3D imaging, and  application of the method to myocardial perfusion SPECT as well as other clinical SPECT imaging applications are important research frontiers. Overall, these results motivate further development of LM-based ASC methods that include data from the scatter window.
\section{Acknowledgements}\label{sec:Ack}
This work was financially supported by NIH R21 EB024647 (Trailblazer award) and by an NVIDIA GPU grant. We also thank the Washington University Center for High Performance Computing for providing computational resources for this project. The center is partially funded by NIH grants 1S10RR022984-01A1 and 1S10OD018091-01.
\bibliographystyle{IEEEbib}
\bibliography{strings,refs}

\begin{thebibliography}{10}

\bibitem{bailey2013evidence}
D.~L. Bailey and K.~P. Willowson,
\newblock ``An evidence-based review of quantitative {SPECT} imaging and
  potential clinical applications,''
\newblock {\em J. Nucl. Med.}, vol. 54, no. 1, pp. 83--89, 2013.

\bibitem{song2011development}
N.~Song, Y.~Du, B.~He, and E.~C. Frey,
\newblock ``Development and evaluation of a model-based downscatter
  compensation method for quantitative {I}-131 {SPECT},''
\newblock {\em Med. Phys.}, vol. 38, no. 6Part1, pp. 3193--3204, 2011.

\bibitem{jha2016no}
A.~K. Jha, B.~Caffo, and E.~C. Frey,
\newblock ``A no-gold-standard technique for objective assessment of
  quantitative nuclear-medicine imaging methods,''
\newblock {\em Phys. Med. Biol.}, vol. 61, no. 7, pp. 2780, 2016.

\bibitem{garcia2007spect}
E.~V. Garcia,
\newblock ``{SPECT} attenuation correction: an essential tool to realize
  nuclear cardiology’s manifest destiny,''
\newblock {\em J. Nucl. Cardiol.}, vol. 14, no. 1, pp. 16, 2007.

\bibitem{hutton2011review}
B.~F. Hutton, I.~Buvat, and F.~J. Beekman,
\newblock ``Review and current status of {SPECT} scatter correction,''
\newblock {\em Phys. Med. Biol.}, vol. 56, no. 14, pp. R85, 2011.

\bibitem{ljungberg2018absolute}
M.~Ljungberg,
\newblock ``Absolute quantitation of {SPECT} studies,''
\newblock in {\em Sem. Nuc. Med.} Elsevier, 2018, vol.~48, pp. 348--358.

\bibitem{king1995attenuation}
M.~A. King, B.~M. Tsui, and T.~Pan,
\newblock ``Attenuation compensation for cardiac single-photon emission
  computed tomographic imaging: Part 1. {I}mpact of attenuation and methods of
  estimating attenuation maps,''
\newblock {\em J. Nucl. Cardiol.}, vol. 2, no. 6, pp. 513--524, 1995.

\bibitem{hashimoto1997scatter}
J.~Hashimoto et~al.,
\newblock ``Scatter and attenuation correction in technetium-99m brain
  {SPECT},''
\newblock {\em J. Nucl. Med.}, vol. 38, no. 1, pp. 157--162, 1997.

\bibitem{mas1990scatter}
J.~Mas, P.~Hannequin, R.~B. Younes, B.~Bellaton, and R.~Bidet,
\newblock ``Scatter correction in planar imaging and {SPECT} by constrained
  factor analysis of dynamic structures (fads),''
\newblock {\em Phys. Med. Biol.}, vol. 35, no. 11, pp. 1451, 1990.

\bibitem{devito1989energy}
R.~P. DeVito, J.~J. Hamill, J.~D. Treffert, and E.~W. Stoub,
\newblock ``Energy-weighted acquisition of scintigraphic images using finite
  spatial filters,''
\newblock {\em J. Nucl. Med.}, vol. 30, no. 12, pp. 2029--2035, 1989.

\bibitem{guerin2010novel}
B.~Gu{\'e}rin and G.~El~Fakhri,
\newblock ``Novel scatter compensation of list-mode {PET} data using spatial
  and energy dependent corrections,''
\newblock {\em IEEE Trans. Med. Imag.}, vol. 30, no. 3, pp. 759--773, 2010.

\bibitem{popescu2006pet}
L.~M. Popescu, R.~M. Lewitt, S.~Matej, and J.~Karp,
\newblock ``{PET} energy-based scatter estimation and image reconstruction with
  energy-dependent corrections,''
\newblock {\em Phys. Med. Biol.}, vol. 51, no. 11, pp. 2919, 2006.

\bibitem{rahman2019fisher}
M.~A. Rahman, Y.~Zhu, E.~Clarkson, M.~A. Kupinski, E.~C. Frey, and A.~K. Jha,
\newblock ``Fisher information analysis of list-mode {SPECT} emission data for
  joint estimation of activity and attenuation distribution,''
\newblock {\em arXiv preprint arXiv:1807.01767}, 2020.

\bibitem{Bouwens:2001}
L.~R. Bouwens, H.~Gifford, R.~V. de~Walle, M.~A. King, I.~Lemahieu, and R.~A.
  Dierckx,
\newblock ``Resolution recovery for list-mode reconstruction in {SPECT},''
\newblock {\em Phys. Med. Biol.}, vol. 46, no. 8, pp. 2239--2253, jul 2001.

\bibitem{caucci2019towards}
L.~Caucci, Z.~Liu, A.~K. Jha, H.~Han, L.~R. Furenlid, and H.~H. Barrett,
\newblock ``Towards continuous-to-continuous 3{D} imaging in the real world,''
\newblock {\em Phys. Med. Biol.}, vol. 64, no. 18, pp. 185007, 2019.

\bibitem{jha2015estimating}
A.~K. Jha and E.~C. Frey,
\newblock ``Estimating {ROI} activity concentration with photon-processing and
  photon-counting {SPECT} imaging systems,''
\newblock in {\em Medical Imaging}. SPIE, 2015, vol. 9412, p. 94120R.

\bibitem{jha2015singular}
A.~K. Jha, H.~H Barrett, E.~C. Frey, E.~Clarkson, L.~Caucci, and M.~A.
  Kupinski,
\newblock ``Singular value decomposition for photon-processing nuclear imaging
  systems and applications for reconstruction and computing null functions,''
\newblock {\em Phys. Med. Biol.}, vol. 60, no. 18, pp. 7359, 2015.

\bibitem{Jha:13:Fully3d}
A.~K. Jha, H.~H. Barrett, E.~Clarkson, L.~Caucci, and M.~A. Kupinski,
\newblock ``Analytic methods for list-mode reconstruction,''
\newblock in {\em Intl Meet Fully Three-Dim Image Recon Rad Nucl Med,
  California}, 2013.

\bibitem{Henscheid:2017}
N.~{Henscheid}, A.~K. {Jha}, and H.~H. {Barrett},
\newblock ``Evaluation of photon processing detectors using the {F}ourier
  crosstalk matrix,''
\newblock in {\em 2017 IEEE Nuclear Science Symposium and Medical Imaging
  Conference (NSS/MIC)}, Oct 2017, pp. 1--4.

\bibitem{kadrmas1997analysis}
D.~J. Kadrmas, E.~C. Frey, and B.~M. Tsui,
\newblock ``Analysis of the reconstructibility and noise properties of
  scattered photons in {T}c-99m {SPECT},''
\newblock {\em Phys. Med. Biol.}, vol. 42, no. 12, pp. 2493, 1997.

\bibitem{parra1998list}
L.~Parra and H.~H. Barrett,
\newblock ``List-mode likelihood: {EM} algorithm and image quality estimation
  demonstrated on 2-{D} {PET},''
\newblock {\em IEEE Trans. Med. Imag.}, vol. 17, no. 2, pp. 228--235, 1998.

\bibitem{jha2013joint}
A.~K. Jha, E.~Clarkson, M.~A. Kupinski, and H.~H. Barrett,
\newblock ``Joint reconstruction of activity and attenuation map using {LM}
  {SPECT} emission data,''
\newblock in {\em Medical Imaging}. SPIE, 2013, vol. 8668, p. 86681W.

\bibitem{hudson1994accelerated}
H.~M. Hudson and R.~S. Larkin,
\newblock ``Accelerated image reconstruction using ordered subsets of
  projection data,''
\newblock {\em IEEE Trans. Med. Imag.}, vol. 13, no. 4, pp. 601--609, 1994.

\bibitem{zubal1994computerized}
I.~G. Zubal, C.~R. Harrell, E.~O. Smith, Z.~Rattner, G.~Gindi, and P.~B.
  Hoffer,
\newblock ``Computerized three-dimensional segmented human anatomy,''
\newblock {\em Med. Phys.}, vol. 21, no. 2, pp. 299--302, 1994.

\end{thebibliography}

\end{document}